\documentclass[usegraphicx,usenatbib]{mn2e}

\def\hcMpc{h^{3}{\rm Mpc}^{-3}}

\title[$UBR$ CCD photometry]
{$UBR$ CCD photometry of Stromlo-APM galaxies}

\author[Jon Loveday]
{Jon Loveday\thanks{E-mail: J.Loveday@sussex.ac.uk}\\
Astronomy Centre, University of Sussex, Falmer, Brighton, BN1 9QJ}

\begin{document}

\maketitle

\begin{abstract}
We present CCD photometry in the Johnson-Kron-Cousins $UBR$ bands for a sample
of 320 galaxies selected from the Stromlo-APM Redshift Survey.
We use this CCD data to estimate the galaxy luminosity function in the 
$U$, $B$, $R$ and $b_J$ bands, finding consistent results with earlier work.
Fainter galaxies serendipitously observed on our CCD frames allow a check 
of the photometric calibration of the APM Galaxy Survey.
We find no evidence of any significant scale error in the APM magnitudes.
\end{abstract}


\section{Introduction}

The Stromlo-APM galaxy survey \citep{lpme96} --- see 
\verb|http://astronomy.sussex.ac.uk/~loveday/sapm| --- 
is a sparse (1:20) sampled
redshift survey containing 1797 galaxies with $b_J < 17.15$
over a very large volume of space.
The solid angle of the survey is 1.3 sr and the median redshift is
about 15,300 km/s.
This makes the survey very well suited to measurements of the galaxy
luminosity function and large-scale clustering.
Recently \citep{love2000}, we obtained near-infrared $K$-band imaging
for a luminosity-selected subsample of 345 galaxies, enabling us to
estimate the $K$-band ($2.2\mu$m) galaxy luminosity function over a range
of 10 mag.

Here we present $UBR$ CCD photometry for the same sample of galaxies
and use this new data to estimate the galaxy luminosity function in 
$U$, $B$, $R$ and $b_J$.

The observations and data reduction are discussed in \S\ref{sec:obs}.
We use these observations to calibrate the APM scans in \S\ref{sec:apmcal}
and present the luminosity functions in \S\ref{sec:lf}.
We conclude in \S\ref{sec:concs}.
Throughout, we assume a Hubble constant of $H_0 = 100$ km/s/Mpc and an
$\Omega_M = 1, \Omega_\Lambda = 0$ cosmology in calculating distances
and luminosities.

\section{Observations and data reduction} \label{sec:obs}

The galaxies observed were the same sample for which we obtained $K$-band
imaging \citep{love2000}.
Briefly, the galaxies were selected to sample absolute $b_J$
magnitude as uniformly as possible.
In this way we were able to estimate the $K$-band luminosity function
over a wide range of luminosities with a sample of modest size.

CCD imaging of the above sample of galaxies was carried out at the
Cerro Tololo Interamerican Observatory (CTIO) 1.5m telescope using
Tek 2048 CCD camera \#5 at f/13.5 (giving a pixel size of $0.24''$)
over the ten nights 1996 September 7--16.
All but the last night were photometric.
Integration times of 120, 300 and 120 s were given in $U$, $B$ and $R$
respectively.
Galaxies were observed at airmass 1--1.3; most were observed with
airmass below 1.2.

The optics of the CTIO 1.5m telescope perform very poorly in the 
$U$ band and our $U$ exposures are of limited use for most galaxies.
However, they were taken under photometric conditions, and so may prove
potentially useful in calibrating deeper $U$-band observations taken under
non-photometric conditions.
Standard star sequences were observed from the compilation of \citet{land92}.
Bias frames and twilight sky flats were taken at the start and end
of each night.

The CCD frames were reduced within the {\sc iraf} environment.
Bias frames were median filtered with minmax rejection to form
a master bias for each night.
The flat-field calibrations through each filter were averaged with cosmic ray
rejection to form master flats for each filter for each night.
Basic reduction was performed using the {\sc quadproc} package, written
at CTIO for reduction of four-amplifier CCD data.
This package trims off the overscan, subtracts the bias frame,
and divides by the master sky flat for the appropriate filter.
Cosmic ray hits were identified and removed with the {\sc cosmicrays} task.
Finally, the $B$ and $U$ images of each field were aligned with the $R$ image
by first estimating
an approximate offset with a single bright star, and then using the
{\sc imalign} task to calculate and apply a more accurate offset.
This alignment technique was also used to co-add multiple observations
of a single field.

\subsection{Image detection and photometry}

We used SExtractor 2.0.15 \citep{ba96}
to detect and measure images in the reduced frames.
In order to determine consistent colours, flux in the $U$, $B$ and $R$ 
bands was measured within the same aperture for any given object.
This aperture was set for each object using the $R$ band image, since this
is the most sensitive band.
For image detection (but not measurement), the $R$ band frames were convolved
with a Gaussian filter of FWHM 5 pixels in order to match the typical
seeing conditions of $\approx 1.2''$, and the detection threshold was set
to 1.5 times the standard deviation of the convolved image.
(For the second night of observations, 1996 Sep 08, we performed $2 \times 2$
on-chip binning, giving a pixel size of $0.48''$.
For this night, we smoothed with a Gaussian of FWHM 3 pixels.)

For both standard stars and galaxies, we used the {\sc mag\_best} estimate
of magnitude.
This yields a pseudo-total magnitude \citep{kron80} except in crowded fields,
when a corrected isophotal magnitude is measured instead.
We used SExtractor's default settings for these adaptive-aperture magnitudes,
and thus measure flux inside an elliptical aperture with semi-major axis
2.5 times larger than the first moment of the light distribution.
This measures roughly 94\% of a galaxy's flux \citep{ba96}; no correction
is applied for the $\sim 6\%$ missing flux, since most other published 
photometry misses a similar fraction of flux.

Magnitude errors were estimated by combining in quadrature SExtractor's
estimate of the error from photon statistics and the difference between
magnitudes measured using local and global estimates of the sky background.
Unfortunately, one of the amplifiers suffered from high read noise, 
making the lower-left quadrant of the CCD image of limited use.
We were careful to avoid placing the target galaxy on or near this bad
quadrant whilst observing, but it did mean that the global background estimate
had to be estimated assuming it to be constant across the entire CCD frame.
Consequently, the global background estimate is a poor approximation to the
local background, and so magnitude errors are overestimated.
Note that the local background estimate was used in determining galaxy 
magnitudes.
This is unaffected by the bad quadrant.

Of the selected sample of 363 galaxies, 320 were observed under photometric
conditions.
Estimated magnitude errors were less than 0.5 mag for 320 observations in $R$,
315 in $B$ and 189 in $U$.

\subsection{Calibration}

We assumed a zero-point of 25 in calculating CCD magnitudes, so that
$m = 25 - 2.5 \lg(\mbox{summed ADU})$.
The following transformation equations were used to convert observed CCD
magnitudes $ubr$ to $UBR$ magnitudes on the Johnson-Kron-Cousins system:
\begin{eqnarray}
R &=& r + r_0 + r_X X + r_c (b-r), \nonumber\\
B &=& b + b_0 + b_X X + b_c (b-r), \label{eqn:trans}\\
U &=& u + u_0 + u_X X + u_c (u-b), \nonumber
\end{eqnarray}
where $r_0, b_0, u_0$ are the zero-point offsets, $r_X, b_X, u_X$ the 
extinction coefficients multiplying the airmass $X$, and $r_c, b_c, u_c$ the
colour terms in the $R, B, U$ bands respectively.

We made a total of 282 standard star observations during the 9 photometric 
nights (an average of 31 per night) and we initially
fitted the transformation parameters independently for each night.
Averaging over nights, with equal weight per night, we obtained:
\begin{eqnarray}
r_0 &=& 1.821 \pm 0.010,\nonumber\\
r_X &=& 0.039 \pm 0.088,\nonumber\\
r_c &=& -0.018 \pm 0.010,\nonumber\\[1mm]
b_0 &=& 2.124 \pm 0.074,\nonumber\\
b_X &=& 0.241 \pm 0.063,\label{eqn:transval}\\
b_c &=& 0.048 \pm 0.009,\nonumber\\[1mm]
u_0 &=& 4.781 \pm 0.624,\nonumber\\
u_X &=& 0.385 \pm 0.533,\nonumber\\
u_c &=& -0.064 \pm 0.037.\nonumber
\end{eqnarray}
Standard star observations with large residuals were omitted from the
fitting procedure, most of these stars fell in the bad quadrant of the CCD.
Since we expect only the extinction parameters to change from night to night,
we held the zero-point and colour terms fixed as given by (\ref{eqn:transval})
and re-fitted the extinction terms for each night.
The nightly extinction terms, along with rms residual magnitude differences
from Landolt's values, are given in Table~\ref{tab:calib}.
Galaxy magnitudes were converted to the Johnson-Kron-Cousins system
using (\ref{eqn:trans}) with zero-point and colour terms from 
(\ref{eqn:transval}) and extinction coefficients from Table~\ref{tab:calib}.
Galaxies observed during non-photometric conditions were rejected.
Photometry for the target galaxies is presented in the Appendix
to this paper.

\begin{table}
 \caption{Standard star extinction coefficients and magnitude residuals.}
 \label{tab:calib}
 \begin{math}
 \begin{array}{lcccccc}
 \hline
 {\rm Night} & r_X & \sigma_r & b_X & \sigma_b & u_X & \sigma_u \\
 \hline
\mbox{1996 Sep 07} & 0.019 & 0.041 & 0.204 & 0.035 & 0.308 & 0.149\\
\mbox{1996 Sep 08} & 0.045 & 0.032 & 0.224 & 0.037 & 0.332 & 0.157\\
\mbox{1996 Sep 09} & 0.024 & 0.027 & 0.200 & 0.026 & 0.389 & 0.150\\
\mbox{1996 Sep 10} & 0.035 & 0.025 & 0.211 & 0.025 & 0.459 & 0.128\\
\mbox{1996 Sep 11} & 0.037 & 0.030 & 0.246 & 0.028 & 0.411 & 0.135\\
\mbox{1996 Sep 12} & 0.050 & 0.030 & 0.272 & 0.023 & 0.448 & 0.193\\
\mbox{1996 Sep 13} & 0.049 & 0.028 & 0.269 & 0.023 & 0.455 & 0.226\\
\mbox{1996 Sep 14} & 0.052 & 0.035 & 0.272 & 0.027 & 0.394 & 0.215\\
\mbox{1996 Sep 15} & 0.053 & 0.022 & 0.259 & 0.022 & 0.356 & 0.200\\
  \hline
 \end{array}
 \end{math}
\end{table}

\subsection{Obtaining $b_J$ magnitudes}

\citet{cn80} have published a transform from $B,R$ magnitudes to the
$B_J, R_F$ system:
\begin{eqnarray}
R_F &=& R - 0.058(B-R) - 0.008,\nonumber\\
(B_J - R_F) &=& -0.027(B-R)^2 + 1.059(B-R) - 0.017.\nonumber\\
\label{eqn:cn}
\end{eqnarray}
However, as pointed out by the referee, the Couch \& Newell $B_J$
passband is rather a poor approximation to the photographic $b_J$ band.

There are several transforms from $B,V$ magnitudes to $b_J$ available in
the literature:
\begin{eqnarray}
b_J &=& B - 0.23(B-V), \label{eqn:kron}\\
b_J &=& B - 0.28(B-V), \label{eqn:bg}\\
b_J &=& B - 0.275(B-V) - 0.067(B-V)^2 + 0.078. \label{eqn:gullixson}
\end{eqnarray}
These equations come from \citet{kron78}, \citet{bg82} and \citet{gbts95}
respectively.

Since we do not have any $V$ band observations, we find a relation between
$B-V$ and $B-R$ by fitting a straight line to these colours using the standard
stars of \citet{land92}.
Excluding one discrepant star, we find that
\begin{equation} \label{eqn:landolt}
B-V = 0.618(B-R) + 0.008,
\end{equation}
with a scatter about this relation of 0.047 magnitudes.

Using the above, we find that the Kron, Blair \& Gilmore
and Gullixson et al. transforms yield consistent $b_J$ magnitudes
(with a scatter $\sim \pm 0.02$ mag), whereas the Couch \& Newell $B_J$
magnitudes are about 0.15 mag fainter.
We use the Gullixson et al. transform (\ref{eqn:gullixson})
for the rest of this paper.

\subsection{Photometric repeatability}

\begin{figure}
\includegraphics[width=\linewidth]{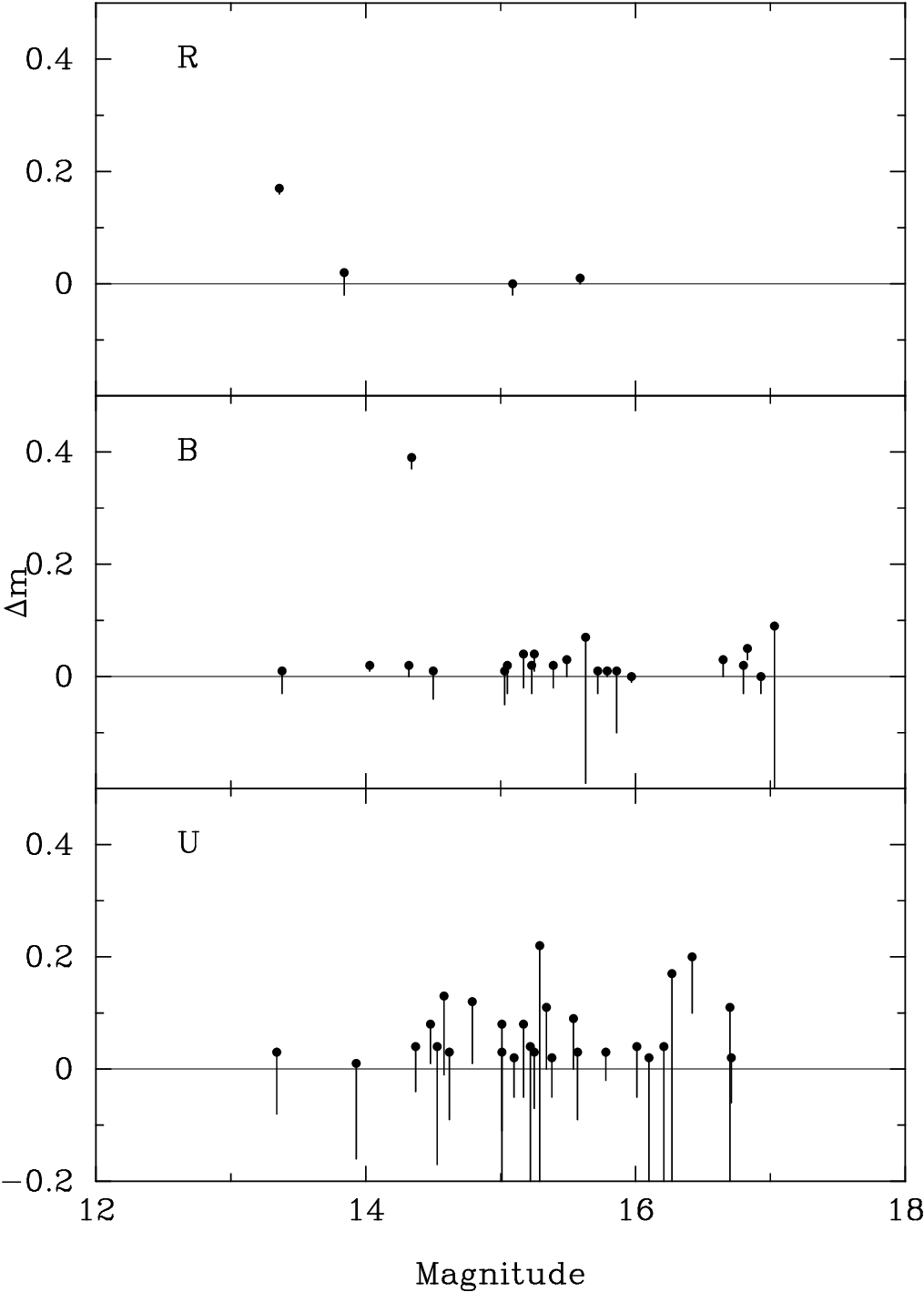}
\caption{Repeatability of galaxy photometry in $R$ (top), $B$ (middle)
and $U$ (bottom).
We plot the rms magnitude between repeated observations
against the magnitude from the co-added data.
The symbols indicate the rms magnitude error and the length of
the line shows the estimated magnitude error.
\label{fig:repeats}}
\end{figure}

Thirty-one galaxies were observed on more than one occasion, 
mostly in $B$ and/or $U$, which allows us to assess the repeatability of
our photometry.
In these cases, we ran SExtractor on the galaxy frames both before
and after co-adding observations.
We obtain final photometry from the co-added frame and use 
the two individual frames and the co-added frame to estimate the rms error.

In Figure~\ref{fig:repeats} we plot both our estimated errors and the rms
errors between repeated observations as a function of magnitude for $R$,
$B$ and $U$.
The location of each symbol indicates the rms magnitude error and the length of
the line shows the estimated magnitude error from the co-added image.
Thus if our estimated errors are a good estimate of the true rms then the lower
ends of the error bars should just reach zero.
For all but one or two galaxies, the estimated error is at least as large
as the rms error.
The points with rms $R$ magnitude error $\approx 0.17$ and rms $B$ 
magnitude error $\approx 0.4$ are from the same galaxy.
This galaxy is in fact one of a close pair of galaxies, and so the large
discrepancy in magnitudes between the two observations presumably reflects
a deblending problem.
Generally however, the error is overestimated.
As discussed above, this is due to the poor global background estimate.
Excluding the galaxy that was poorly deblended from a close neighbour,
our repeated observations lead us to estimate typical magnitude errors
for our target galaxies
in the $R$, $B$ and $U$ bands of $0.010 \pm 0.006$, $0.025 \pm 0.005$
and $0.070 \pm 0.011$ respectively.

\begin{figure}
\includegraphics[width=\linewidth]{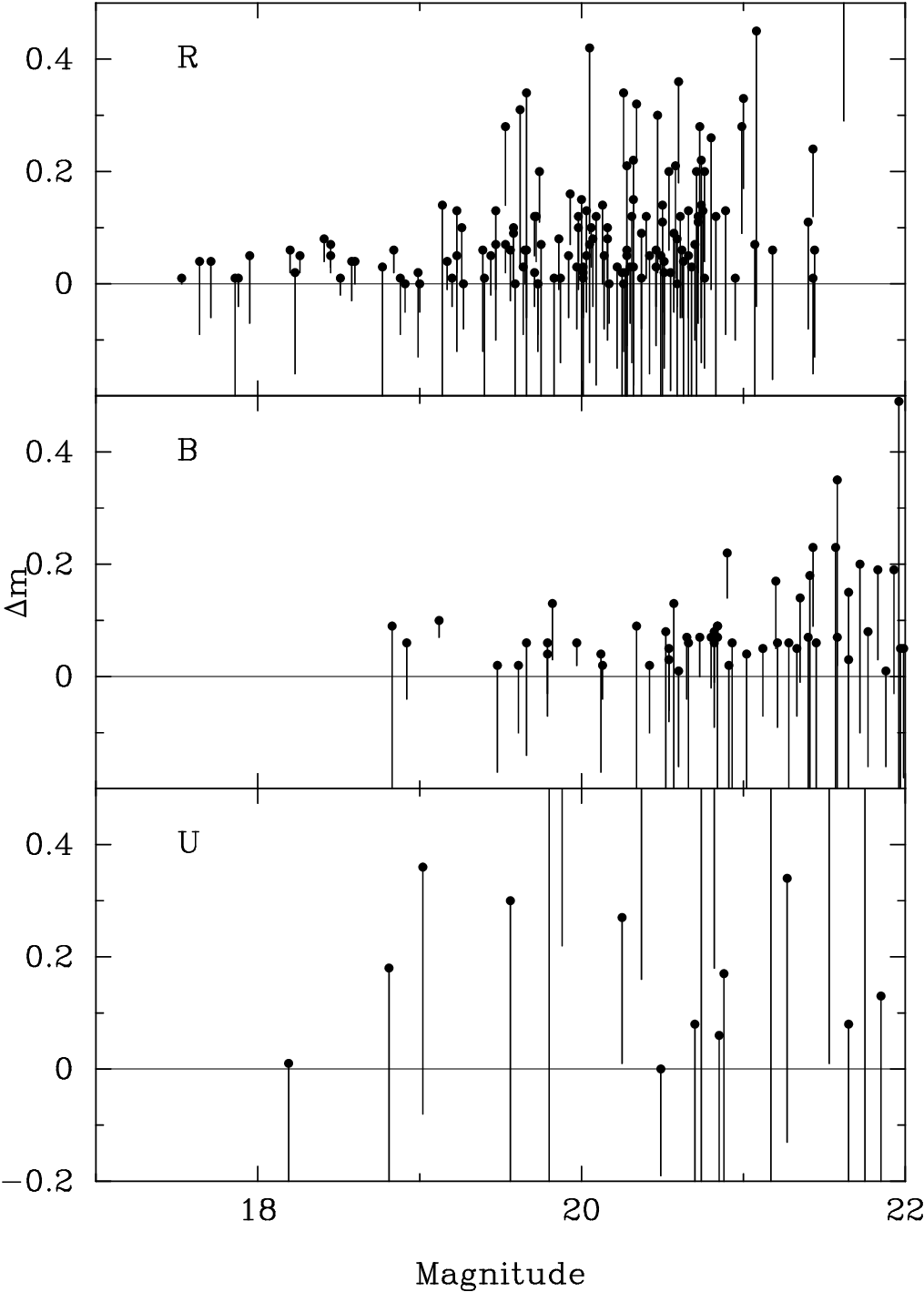}
\caption{Repeatability of faint galaxy photometry in $R$ (top), $B$ (middle)
and $U$ (bottom).
As Figure~\ref{fig:repeats}, but for all reliable, multiply-detected galaxies,
not just target galaxies.
\label{fig:repeat_all}}
\end{figure}

We have also investigated the photometric reliability of fainter galaxies
which by chance also lie on our CCD frames.
Four of our targets were observed in all 3 bands on more than one occasion,
allowing us to measure magnitude differences in each band as a function
of magnitude.
The results are shown in Figure~\ref{fig:repeat_all}.
We see that the majority of $R$ and $B$ magnitudes have errors $\la 0.1$
mag down to 21st magnitude.
Most galaxies with a repeat error larger than this have their magnitude
error set appropriately.
The $U$ band photometry, however, is very unreliable below 18th magnitude
(indeed very few sources fainter than this are reliably detected in both
$U$ exposures).

We have checked that we are not systematically missing $B$-band flux
by comparing flux measured in co-added $B$ frames with the flux
measured in individual frames.  
No systematic trend is seen for $B < 22$.

In conclusion, the $R$ and $B$ band photometry appears to be reliable to
about 21st magnitude.
$U$ band photometry is not useful below 18th magnitude.

\section{Calibrating APM Scans}  \label{sec:apmcal}

\begin{figure}
\includegraphics[angle=-90,width=\linewidth]{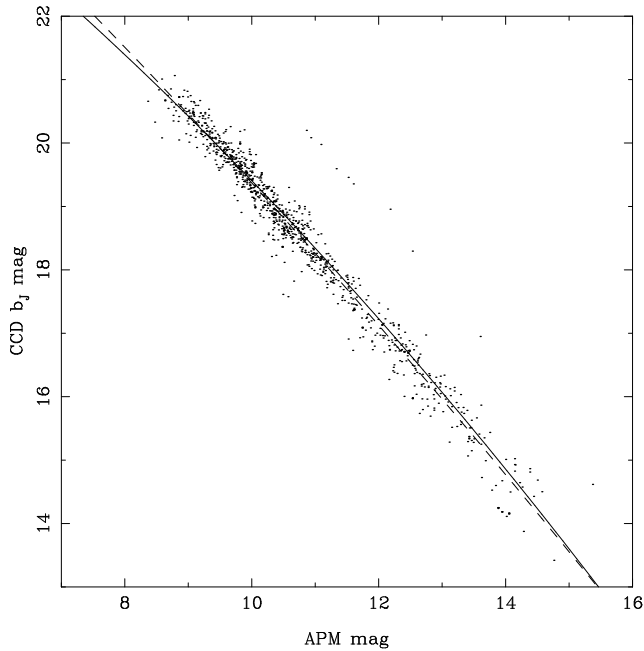}
\caption{CCD $b_J$ magnitude plotted against APM magnitude.
Lines indicate quadratic functions used to convert from APM magnitudes
to $b_J$ magnitudes, allowing for photographic saturation.
The continuous line shows the conversion assumed in the original APM
survey \citep{mselp90} and the dashed line the best fit to our CTIO data.
\label{fig:apmcal}}
\end{figure}

Since the CCD frames cover an area of sky around 8 arcmin on a side,
they provide $U, B, R$ photometry for
many objects in each field in addition to the target galaxy.
We therefore matched the images detected by SExtractor with images in the
APM scans of \citet{msel90}.
Using the matched objects in each frame, we calculated a 6-parameter transform 
from CCD pixel coordinates to APM plate coordinates and thence to RA \& Dec.

In Figure~\ref{fig:apmcal} we plot the $b_J$ magnitude derived from
our CCD data using the \citet{gbts95} transform
against the matched APM magnitude.
This plot compares magnitudes for 1067 galaxies, ie. sources classified
as non-stellar in the APM scans {\em and} with an SExtractor stellar
likelihood {\sc class\_star} $< 0.5$.
We have excluded sources detected in the lower-left quadrant of the CCD due
to the excessive read noise in this amplifier of the detector, all
sources with any of the SExtractor warning flags set \citep{ba96}, and
any sources with an estimated $b_J$ magnitude error (obtained by adding 
in quadrature estimated errors in $B$ and $R$) greater than 0.1 mag.
All sources shown are thus expected to have ``clean'' CCD photometry.

The APM magnitudes have been corrected for vignetting and differential
desensitisation and have been matched to be consistent from plate-to-plate, 
using the procedures described in \citet{mes90}, but are uncorrected for
photographic non-linearities and are opposite in sign from standard magnitudes,
ie. a large APM magnitude implies a large flux.
In Figure~\ref{fig:apmcal} the continuous line shows the quadratic
calibration used by \citet{mselp90}, viz.
\begin{equation}
b_J = 27.405  - 0.5588 m_{apm} - 0.0241 m_{apm}^2,
\end{equation}
and the dashed line shows the best fit quadratic to the present CCD data,
\begin{equation}
b_J = 28.984  - 0.8263 m_{apm} - 0.013 m_{apm}^2.
\end{equation}
It is clear that the two calibrations agree well over 
the entire magnitude range plotted, although there is an overall
offset of $-0.052 \pm 0.009$ between the new and old CCD calibration, with
only a very weak magnitude dependence.

There are ten outliers above and to the right of the main locus of
points.
These galaxy images have been visually inspected and most correspond
to galaxies with close companions (either galaxies or stars), and so presumably
in these cases the APM scans have blended in flux from the companion object
and so have overestimated the APM magnitude.

Overall, our CCD photometry is in good agreement with the previous calibration
of APM galaxies.
There is no evidence for an $\approx$ 20\% scale error in APM magnitudes
as seen by \citet{mfs95} in one of their fields, and so presumably it is
just the field in question (GSM) which suffers from this problem.
Our calibrations are consistent with those recently presented by
\citet{norberg2002}.
The unusually steep number counts seen by \citet{mselp90} are thus not 
explained by magnitude scale errors, but are likely to be due to a combination
of large scale structure and low-redshift evolution.

\section{Luminosity functions} \label{sec:lf}

In this section we estimate luminosity functions in the $R$, $B$, $U$ and $b_J$
bands.
Estimates of the LF in these bands over a smaller area of sky have been 
presented by \citet{mrsf98}, who used CCD calibrated photographic plates.
Here we present LFs in these bands directly from CCD photometry.

In calculating rest-frame absolute magnitudes from the CCD photometry
we use the $k$-corrections of \citet{fg94}, who
tabulate colours and $k$-corrections for four different
Hubble types (E, Sbc, Scd and Im) at redshifts $z = 0.0$, 0.1, 0.2, 0.4 
and 0.6.
For each galaxy in our sample, we interpolate the Frei and Gunn colours
to the observed redshift $z$ and then find which of the four Hubble types
provides the closest match to our observed $(B-R)$ colour.
We then use this Hubble type to estimate the $k$-correction in the
appropriate band at redshift $z$.

Since the sample was selected using APM $b_J$ magnitude, it is incomplete
in the CCD $R$, $B$, $U$ and $b_J$ bands.
Following \citet{love2000}, we estimate a bivariate luminosity function (BLF)
$\phi(M_1, M_2)$, where $M_1$ is the absolute magnitude inferred from the
APM data and $M_2$ is the CCD band being considered.
We use a density-independent stepwise maximum likelihood (SWML) method
to estimate $\phi(M_1, M_2)$, allowing for the known selection in $b_J$ 
flux and $M_1$ absolute magnitude.
As with all density-independent estimators, information about the overall
normalization is lost.
We therefore normalize our BLF to the mean density of galaxies with 
$-22 \le M_1 \le -13$ in the full Stromlo-APM sample, 
$\bar{n} = 0.071 h^3{\rm Mpc}^{-3}$, calculated as described by 
\citet{lpem92}\footnote{
Note that the density $\bar{n} = 0.047 h^3{\rm Mpc}^{-3}$ quoted by 
Loveday et al.\ (1992) is for the restricted magnitude range 
$-22 \le M_B \le -15$.}.
We then integrate $\phi(M_1, M_2)$ over $M_1$ to obtain $\phi(M_2)$.
See \citet{love2000} for full details of the procedure, which is carried out
in turn using the $R$, $B$, $U$ CCD magnitudes and also using a $b_J$ magnitude
inferred from the $R$ and $B$ magnitudes using (\ref{eqn:gullixson}) and
(\ref{eqn:landolt}).

\begin{figure}
\includegraphics[angle=-90,width=\linewidth]{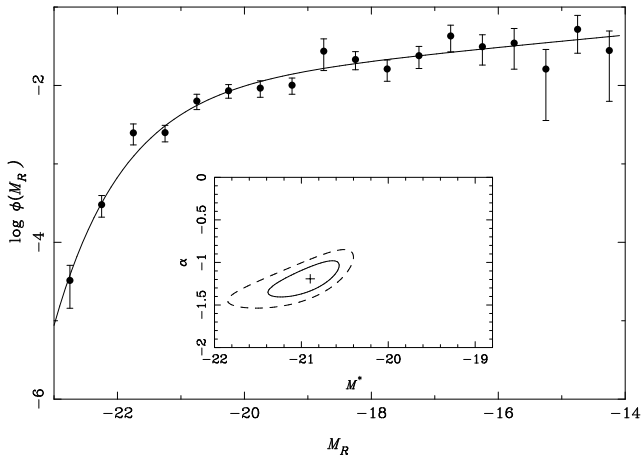}
\caption{The $R$ band luminosity function estimated from our sample
(symbols) together with the best-fit Schechter function.
The inset shows the 1 and 2 $\sigma$ likelihood contours for the shape 
parameters $\alpha$ and $M^*$.
\label{fig:lumR}}
\end{figure}

\begin{figure}
\includegraphics[angle=-90,width=\linewidth]{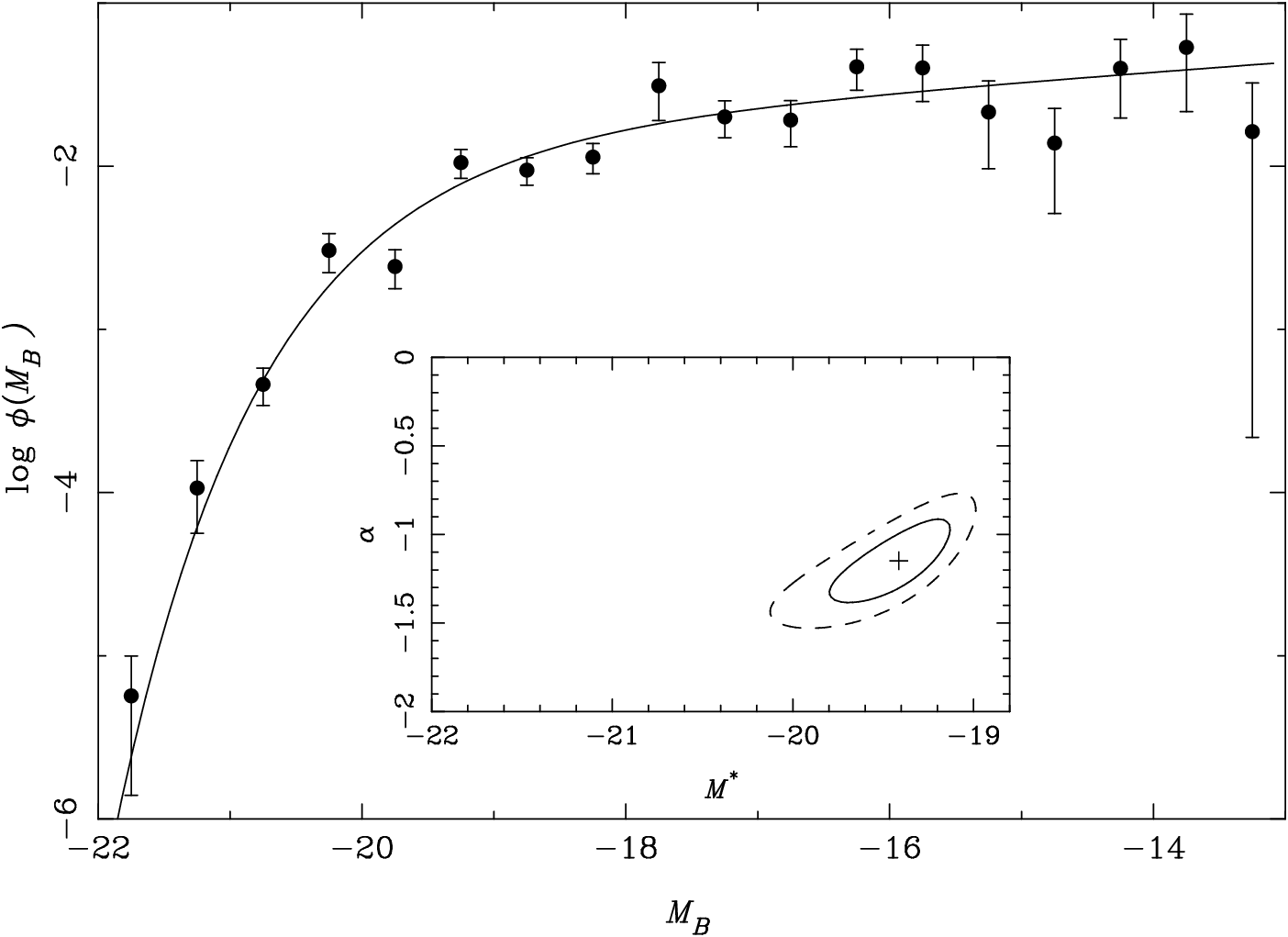}
\caption{The $B$ band luminosity function estimated from our sample
(symbols) together with the best-fit Schechter function.
The inset shows the 1 and 2 $\sigma$ likelihood contours for the shape 
parameters $\alpha$ and $M^*$.
\label{fig:lumB}}
\end{figure}

\begin{figure}
\includegraphics[angle=-90,width=\linewidth]{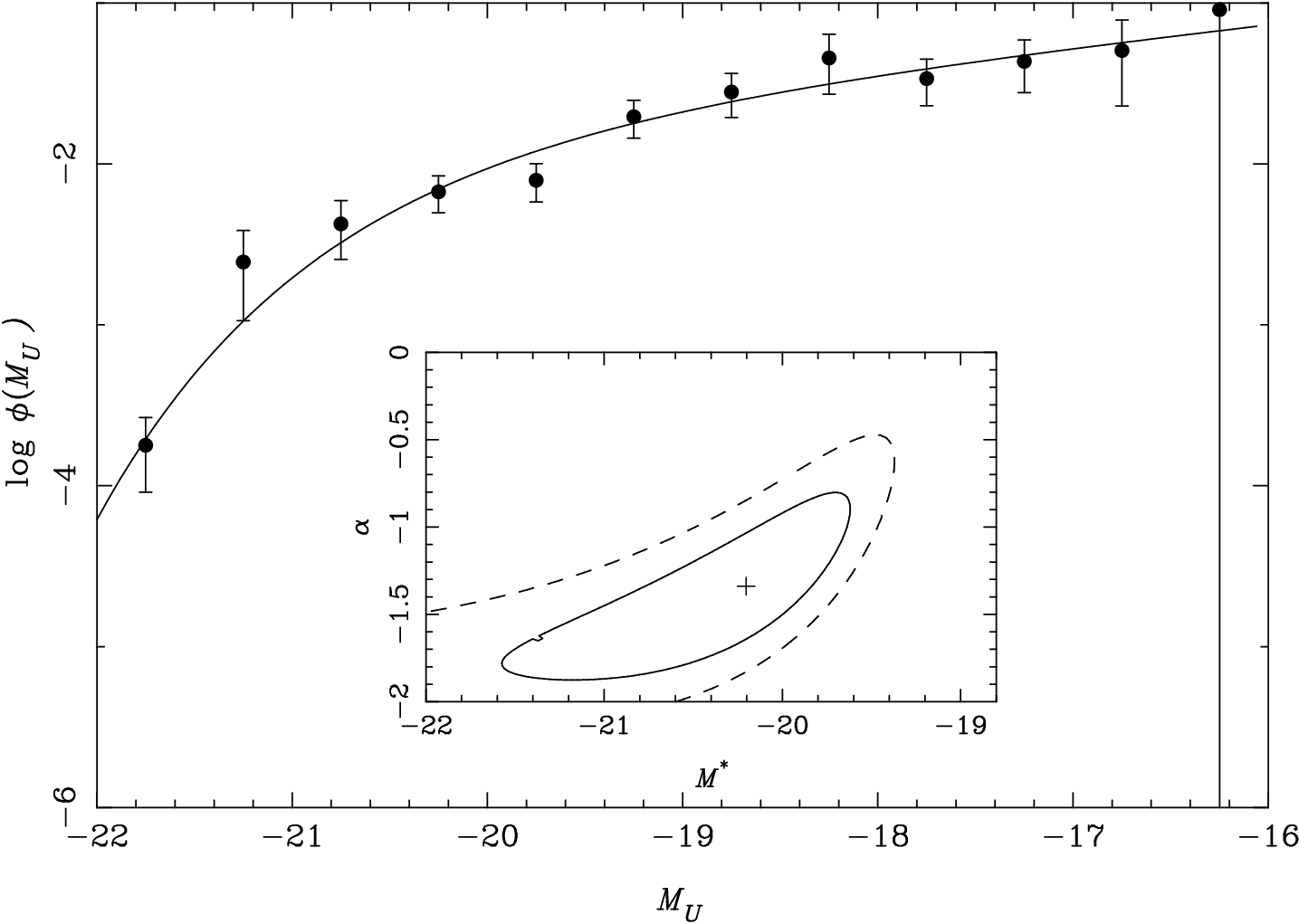}
\caption{The $U$ band luminosity function estimated from our sample
(symbols) together with the best-fit Schechter function.
The inset shows the 1 and 2 $\sigma$ likelihood contours for the shape 
parameters $\alpha$ and $M^*$.
\label{fig:lumU}}
\end{figure}

\begin{figure}
\includegraphics[angle=-90,width=\linewidth]{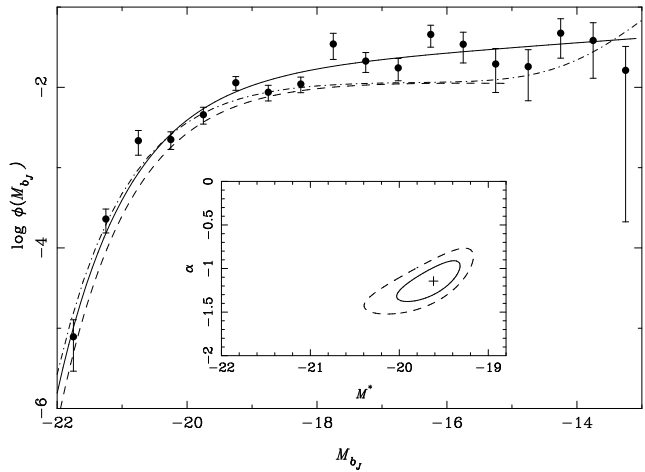}
\caption{The $b_J$ band luminosity function estimated from our sample
(symbols) together with the best-fit Schechter function.
The dashed line shows the Schechter function fit from \citet{lpem92} and
the dot-dashed line the double power-law fit from \citet{love97}.
The inset shows the 1 and 2 $\sigma$ likelihood contours for the shape 
parameters $\alpha$ and $M^*$.
\label{fig:lumBj}}
\end{figure}

\begin{table}
 \begin{center}
 \caption{Schechter function fits to luminosity functions.}
 \label{tab:schec}
 \begin{math}
 \begin{array}{lccc}
 \hline
 {\rm Band} & \alpha & M^* & \phi^* (\hcMpc)\\
 \hline
{\bmath R} & -1.19 \pm 0.20 & -20.90 \pm 0.42 & 0.014 \pm 0.011\\
R^1 & (-1.2) & -21.24 \pm 0.10 & 0.007 \pm 0.003\\
{r^*}^2 & -1.15 \pm 0.03 & -20.67 \pm 0.03 & 0.019 \pm 0.002\\
r^3 & -0.70 \pm 0.05 & -20.29 \pm 0.02 & 0.019 \pm 0.001\\
{\bmath B} & -1.15 \pm 0.25 & -19.41 \pm 0.35 & 0.020 \pm 0.013\\
B^1 & -1.20 \pm 0.12 & -19.75 \pm 0.14 & 0.007 \pm 0.003\\
B^4 & -1.12 \pm 0.05 & -19.43 \pm 0.06 & 0.013 \pm 0.002\\
{\bmath U} & -1.34 \pm 0.55 & -20.20 \pm 1.00 & 0.022 \pm 0.050\\
U^1 & (-1.2) & -19.74 \pm 0.06 & 0.007 \pm 0.003\\
{u^*}^2 & -1.31 \pm 0.09 & -18.24 \pm 0.07 & 0.047 \pm 0.011\\
{\bmath b_J}& -1.14 \pm 0.25 & -19.62 \pm 0.35 & 0.019 \pm 0.013\\
b_J^5 & -0.97 \pm 0.15 & -19.50 \pm 0.13 & 0.014 \pm 0.002\\
b_J^6 & -1.19 \pm 0.01 & -19.79 \pm 0.04 & 0.016 \pm 0.001\\
  \hline
 \end{array}
 \end{math}
\end{center}
Notes: 
Quantities in parentheses were held fixed.
Bands designated in bold face are from the present work.  Other are from:\\
$^1$ \citet{mrsf98}.\\
$^2$ \citet{blan2001} ($\Omega_m = 1$).\\
$^3$ \citet{lin96a}.\\
$^4$ \citet{mcpwg98}.\\
$^5$ \citet{lpem92}.\\
$^6$ \citet{madg2002}.
\end{table}

Our estimated luminosity functions are shown in 
Figures~\ref{fig:lumR}--\ref{fig:lumBj}.
The curves show \citet{schec76} function fits to the SWML estimates using least
squares, since we are unable to fit parametric models directly to incomplete
data.
We allow for finite bin width by calculating the mean of the predicted 
$\langle \phi_j \rangle$ at the absolute magnitude of each galaxy in each bin, 
rather than simply calculating $\phi(M)$ at the bin centre.
The insets show 1 and 2 $\sigma$ likelihood contours for the shape parameters
$\alpha$ and $M^*$, where the normalization $\phi^*$ is adjusted to maximize 
the likelihood at each grid point.
The best-fit Schechter parameters in each band are given in 
Table~\ref{tab:schec}, along with other estimates from the literature.
The quoted errors on $\alpha$ and $M^*$ come from the bounding-box of the 
1$\sigma$ likelihood contours.
Since our sample was $b_J$-selected, our galaxies tend to be fainter in
the $U$ and $R$ bands, and so we sample the faint end of the luminosity 
function in these bands better than the bright end.
The LF estimates were normalized to the same number density as the full 
Stromlo-APM sample, and hence the errors in $\phi^*$
are dominated by the uncertainty in shape of the LFs.

\subsection{$R$-band luminosity function}

Our estimate (Figure~\ref{fig:lumR})
is consistent in shape with that of \citet{mrsf98}, who held the faint-end slope
fixed and found a characteristic magnitude rather brighter than ours,
but in agreement within the quoted errors.
Their $\phi^*$ is much lower (by a factor of more than two) than
other estimates shown in Table~\ref{tab:schec}, reflecting the unusually
low normalization of the Durham-AAT survey, eg. \citet{ecbhg96}.

We are also in good agreement with the LF measured in $r^*$ from
the Sloan Digital Sky Survey \citep[SDSS,][]{blan2001}, 
albeit with considerably larger uncertainties than this much larger sample.

In agreement with the above authors, we measure a significantly steeper 
faint-end slope and brighter characteristic magnitude
than was measured from the Las Campanas Redshift Survey 
\citep{lin96a}.
As discussed by \citet{blan2001}, this difference is most likely due to the 
fact that \citet{lin96a} measured isophotal magnitudes corresponding
to the relatively bright isophote $\mu_R \approx 23$ mag arcsec$^{-2}$
and that they excluded galaxies of low central surface brightness.

\subsection{$B$-band luminosity function}

Our estimate (Figure~\ref{fig:lumB})
is in good agreement with previous measurements from
the Second Southern Sky Redshift Survey (SSRS2) by \citet{mcpwg98}
and by \citet{mrsf98}.
Once again the low normalization of the Durham-AAT survey stands out.

\subsection{$U$ band luminosity function}

Our $U$-band LF is shown in Figure~\ref{fig:lumU}.
By convolving a Schechter luminosity function with a Gaussian of width 0.07
magnitudes, we have verified that the rms magnitude error of 0.07 mag in
our $U$-band photometry has a negligible effect on the fitted Schechter
function parameters.
Given the large uncertainty in our estimate of the characteristic magnitude
in this band, we are consistent with \citet{mrsf98}.
We are apparently inconsistent with the much fainter $M^*$ found in the SDSS
$u^*$ band by \citet{blan2001}.
Note that the faint-end slope $\alpha$ is consistent between our estimates,
and so the $\alpha$-$M^*$ correlation is not responsible for the differing
estimates of the characteristic magnitude $M^*$.

Part of this discrepancy can be explained by the differing
response function of SDSS $u^*$ compared with Johnson $U$.
In order to derive an approximate conversion from Johnson $U$ to SDSS $u^*$,
we start with the following transform equations from \citet{smith2002}:
\begin{eqnarray*}
u^* - g^* &=& 1.33(U-B) + 1.12,\\
g^* &=& V + 0.54(B-V) - 0.07.
\end{eqnarray*}
Eliminating $g^*$,
\begin{equation} \label{eqn:u}
u^* = 1.33(U-B) + 0.54(B-V) + V + 1.05.
\end{equation}
Substituting (\ref{eqn:landolt}) into (\ref{eqn:u}), we obtain
\[
B - u^* = 0.29(B-R) - 1.33(U-B) - 1.05.
\]
Using this equation to convert our $UBR$ magnitudes to $u^*$, and fitting
a straight line to $u^*$ vs $U$, we find
\begin{equation} \label{eqn:uu}
u^* \approx 1.016 U + 0.27, 
\end{equation}
with a scatter of 0.25 magnitudes.
Our typical $U$ magnitude of $U \approx 16$ thus corresponds to 
$u^* \approx 16.5$, which helps to explain some, but not all, of the
discrepancy between our estimate
of $M^*$ in the $U$ band and the \citet{blan2001} estimate of $M^*$ in
the SDSS $u^*$ band.

The discrepancy might be further reduced by the fact that 
the $U$ response is slightly bluer than $u^*$. 
Thus the most luminous galaxies in $U$, which 
dominate our $M^*$ estimate and which will be
intrinsically blue in a $b_J$-selected survey, will be systematically fainter
in $u^*$.
This effect is not accounted for in our simple conversion from $U$ to $u^*$
(\ref{eqn:uu}).

\subsection{$b_J$ luminosity function}

Our estimate (Figure~\ref{fig:lumBj}) agrees well with that measured 
using APM magnitudes from the full Stromlo-APM survey \citep{lpem92},
although there is evidence for a slightly steeper faint-end slope $\alpha$ 
in the present result.
Our estimate of the characteristic magnitude $M^*$ is also slightly brighter.
This is due to the use of the \citet{gbts95} $b_J$ transform used here,
whereas \citet{lpem92} used the \citet{cn80} transform, which yields
$B_J$ magnitudes fainter by about 0.15 mag.

From the 2dF Galaxy Redshift Survey, \citet{madg2002} find a slightly
brighter characteristic magnitude than our result.
This is likely to be due to the significantly deeper mean redshift of the 
2dF survey compared with the Stromlo-APM survey.
We are not able to probe far enough down the faint end of the LF to confirm
or refute the upturn in the LF around $b_J \approx -14$ claimed by 
\citet{love97}.

\section{Conclusions} \label{sec:concs}

We have presented CCD photometry in the Johnson-Kron-Cousins $UBR$ bands 
for a sample of 320 galaxies selected from the Stromlo-APM Redshift Survey.
We have derived $b_J$ magnitudes from this data in order to check 
the calibration of the APM scans.
We find no evidence for a significant scale error in the APM 
magnitudes, in agreement with \citet{norberg2002}, but contrary to
the findings of \citet{mfs95}.
It thus appears likely that the apparent scale error found by Metcalfe et al.
was due to a single bad field.

We have measured the galaxy luminosity function in the 
$U$, $B$, $R$ and $b_J$ bands.
The $U$-band LF is consistent with the previous determination from calibrated
photographic plates by \citet{mrsf98}, but 
inconsistent with the $u^*$ LF measured from the Sloan Digital
Sky Survey \citep{blan2001}.
The discrepancy is lessened, however, once the difference between Johnson $U$
and SDSS $u^*$ bands is allowed for.
The $B$, $R$ and $b_J$ band LFs are also consistent with previous 
determinations, in particular with the $b_J$ luminosity function determined
from photographic magnitudes in the full Stromlo-APM survey \citep{lpem92}.

The original aims of this project included estimating bivariate luminosity
functions and subdividing the sample by restframe $U-R$ colour in order to
compare with Canada France Redshift Survey (CFRS) observations 
\citep{lilly95} at higher 
redshifts and so to constrain galaxy evolution models.
The limited depth of our $U$ band data precludes the latter, and has now
been superseded by the Sloan Digital Sky Survey \citep[SDSS,][]{york2000},
which has already obtained high quality five-colour photometry for 
hundreds of thousands of galaxies.
We plan to make use of SDSS data for such investigations.

\section*{Acknowledgments}

It is a pleasure to thank the CTIO staff for their excellent support,
Simon Lilly for sharing the observing, Steve Maddox and Nigel Metcalfe 
for useful discussions and the referee for a very constructive report.

{}

\appendix

\section{The CCD data}

In Table~\ref{tab:the_cat} we present a sampling of our CCD data (42 galaxies
out of a total of 320).
We have included galaxies even with very large ($\ga 1$ mag)
colour errors; the user may wish to treat these as non-detections.
The complete catalogue will be available from the VizieR Catalogue Service
(http://vizier.u-strasbg.fr/).
The first five columns of this table come from the Stromlo-APM survey
(Loveday et al.\ 1996).
The subsequent seven columns are derived from our new observations.
Each column in the table is described below.
\begin{description}

\item[(1) Name:]
Galaxy naming follows the same convention as the APM Bright Galaxy Catalogue
\citep{love96} and the Stromlo-APM Redshift Survey (Loveday et al.\ 1996).

\item[(2), (3) RA, dec:]
Right ascension (hours, minutes, seconds) and declination (degrees, arcminutes,
arcseconds) in 1950 coordinates.

\item[(4) $b_J$:]
$b_J$ magnitude.

\item[(5) $cz$:]
Heliocentric recession velocity in km/s.

\item[(6) $R$:]
$R$ magnitude and its estimated error.

\item[(7) $B-R$:]
$B-R$ colour and its estimated error.

\item[(8) $U-B$:]
$U-B$ colour and its estimated error.

\item[(9), (10) Maj, Min:]
Semi-major and minor axes of measurement ellipse in arcseconds.

\item[(11) PA:]
Position angle in degrees measured clockwise from south-north line.

\item[(12) Flags:]
Flags output by SExtractor \citep{ba96}.

\end{description}

\begin{table*}
\begin{minipage}{\textwidth}
 \caption{CCD galaxy photometry.}
 \label{tab:the_cat}
 \begin{tt}
 \begin{tabular}{rrrrrrrrrrrr}
  \hline
  \hline
   \multicolumn{1}{c}{\rm Name} & \multicolumn{1}{c}{\rm RA} & 
   \multicolumn{1}{c}{\rm Dec}
   & $b_J$ & $cz$ & \multicolumn{1}{c}{$R$} & \multicolumn{1}{c}{$B-R$} 
   & \multicolumn{1}{c}{$U-B$} & {\rm Maj} & {\rm Min} & {\rm PA} & {\rm Flags}\\
   \hline
   075+069-077 & 21 26 16.37 & -68 39 53.3 & 16.32 & 11033 & 15.09 $\pm$ 0.04 &  1.49 $\pm$ 0.08 &  0.27 $\pm$ 0.18 &  19 &  12 &  52 &  2 \\
076-113-015 & 22 56 20.64 & -69 43 06.6 & 16.48 &  3813 & 15.63 $\pm$ 0.18 &  1.03 $\pm$ 0.24 & -0.94 $\pm$ 0.22 &  32 &  14 &  60 &  0 \\
077+055+032 & 23 11 33.04 & -70 38 10.8 & 17.05 & 33310 & 16.44 $\pm$ 0.01 & -1.44 $\pm$ 0.01 & -0.73 $\pm$ 0.06 &  16 &   9 & 116 &  3 \\
077+062-116 & 23 11 36.12 & -67 52 12.9 & 16.57 & 29923 & 14.89 $\pm$ 0.04 & -0.94 $\pm$ 0.09 &  0.00 $\pm$ 0.00 &  28 &  22 & 159 &  3 \\
078-130-118 & 00 21 06.17 & -62 46 46.8 & 16.52 & 12128 & 16.43 $\pm$ 0.01 &  0.47 $\pm$ 0.01 & -3.93 $\pm$ 0.11 &  15 &   9 & 107 &  3 \\
078+012+091 & 23 57 37.39 & -66 47 31.6 & 15.42 & 21754 & 14.00 $\pm$ 0.14 &  1.86 $\pm$ 0.18 & -0.77 $\pm$ 0.21 &  28 &  25 &  66 &  2 \\
078+109+066 & 23 39 46.85 & -66 14 01.5 & 14.62 & 10150 & 13.61 $\pm$ 0.01 &  0.99 $\pm$ 0.02 &  0.00 $\pm$ 0.00 &  48 &  21 &  84 &  3 \\
080-018-033 & 01 31 33.11 & -64 28 28.6 & 17.01 &  8088 & 16.42 $\pm$ 0.33 &  1.05 $\pm$ 0.49 & -0.66 $\pm$ 0.75 &  17 &  14 & 127 &  0 \\
080+009+024 & 01 26 43.01 & -65 31 49.0 & 14.84 &  1624 & 13.68 $\pm$ 0.02 &  1.08 $\pm$ 0.02 & -0.73 $\pm$ 0.02 &  26 &  21 &  14 &  0 \\
082+032-078 & 02 50 37.37 & -63 38 30.4 & 16.68 & 30356 & 15.73 $\pm$ 0.03 &  1.29 $\pm$ 0.06 & -0.16 $\pm$ 0.45 &  22 &   9 & 102 &  0 \\
107-053-114 & 21 24 40.36 & -62 56 09.3 & 17.09 &  8507 & 16.23 $\pm$ 0.02 &  0.89 $\pm$ 0.10 & -0.67 $\pm$ 0.21 &  24 &   8 & 152 &  0 \\
108+093+006 & 21 43 24.76 & -65 10 25.7 & 16.03 & 10497 & 15.29 $\pm$ 0.01 &  0.95 $\pm$ 0.07 & -0.42 $\pm$ 0.17 &  21 &  14 &  38 &  0 \\
108-114+028 & 22 20 24.21 & -65 32 46.0 & 16.63 &  6146 & 15.77 $\pm$ 0.09 &  0.97 $\pm$ 0.18 &  0.00 $\pm$ 0.00 &  28 &  16 &  99 &  0 \\
108-105+055 & 22 19 19.57 & -66 03 24.1 & 14.56 & 10779 & 13.58 $\pm$ 0.05 &  1.25 $\pm$ 0.06 &  0.00 $\pm$ 0.00 &  54 &  22 &  49 &  2 \\
108-083-130 & 22 13 20.13 & -62 38 40.3 & 16.85 & 36934 & 15.66 $\pm$ 0.14 &  1.41 $\pm$ 0.17 & -1.15 $\pm$ 0.54 &  17 &  15 & 131 &  0 \\
109+063+033 & 22 32 40.84 & -65 41 37.3 & 16.69 & 21798 & 14.91 $\pm$ 0.02 &  1.84 $\pm$ 0.07 &  0.00 $\pm$ 0.00 &  19 &  15 & 128 &  0 \\
109+014+028 & 22 41 34.01 & -65 37 16.2 & 14.84 &  3269 & 14.02 $\pm$ 0.03 &  1.16 $\pm$ 0.04 & -0.71 $\pm$ 0.08 &  26 &  26 &  63 &  2 \\
110+020+024 & 23 24 18.95 & -65 32 45.2 & 15.04 &  1991 & 13.23 $\pm$ 0.06 &  0.88 $\pm$ 0.07 & -1.37 $\pm$ 0.11 &  79 &  48 & 164 &  2 \\
111+063-007 & 23 50 48.93 & -59 58 26.5 & 14.81 &  3320 & 14.12 $\pm$ 0.02 &  0.94 $\pm$ 0.04 & -0.54 $\pm$ 0.14 &  37 &  17 &  32 &  0 \\
111+118+005 & 23 42 33.53 & -60 08 39.0 & 15.52 &  3414 & 14.90 $\pm$ 0.10 &  0.96 $\pm$ 0.11 & -0.69 $\pm$ 0.13 &  65 &  15 &  72 &  0 \\
111+025+058 & 23 56 29.36 & -61 11 27.6 & 16.20 & 28872 & 15.14 $\pm$ 0.05 &  1.76 $\pm$ 0.14 &  0.00 $\pm$ 0.00 &  28 &  16 & 173 &  2 \\
112+092-034 & 00 24 46.45 & -59 24 30.9 & 15.97 & 11668 & 14.71 $\pm$ 0.09 &  1.34 $\pm$ 0.09 & -0.18 $\pm$ 0.26 &  23 &  17 &  77 &  0 \\
112-083-068 & 00 50 12.34 & -58 47 36.5 & 16.15 &  5150 & 19.43 $\pm$ 0.15 & -1.02 $\pm$ 0.60 & -0.62 $\pm$ 1.13 &   5 &   4 &  88 &  3 \\
112+079+009 & 00 26 25.33 & -60 13 13.5 & 16.62 &  4720 & 18.72 $\pm$ 0.07 &  1.56 $\pm$ 0.14 &  0.09 $\pm$ 0.55 &   3 &   3 &  17 &  0 \\
113+004+088 & 01 15 15.04 & -61 43 42.3 & 15.31 &  8591 & 14.16 $\pm$ 0.01 &  1.31 $\pm$ 0.01 & -0.45 $\pm$ 0.08 &  34 &  12 &   6 &  2 \\
114-031+009 & 01 58 39.99 & -60 14 34.1 & 17.06 &  6743 & 15.85 $\pm$ 0.07 &  1.28 $\pm$ 0.09 & -0.63 $\pm$ 0.33 &  25 &   8 & 122 &  0 \\
114-015+074 & 01 56 18.04 & -61 27 21.5 & 14.80 &  7002 & 12.52 $\pm$ 0.05 &  1.38 $\pm$ 0.05 & -0.29 $\pm$ 0.15 &  64 &  48 & 108 &  0 \\
114-123-003 & 02 12 22.81 & -59 56 14.5 & 16.35 &  1471 & 15.95 $\pm$ 0.28 &  0.74 $\pm$ 0.32 & -0.36 $\pm$ 0.25 &  57 &  11 &  99 &  2 \\
116-048-060 & 03 17 06.64 & -58 58 03.9 & 16.62 & 21466 & 15.38 $\pm$ 0.05 &  1.24 $\pm$ 0.05 & -0.38 $\pm$ 0.19 &  14 &  13 &  89 &  0 \\
117+080-131 & 03 36 49.50 & -57 36 19.7 & 16.02 &  4952 & 14.73 $\pm$ 0.01 &  1.24 $\pm$ 0.01 & -0.43 $\pm$ 0.09 &  23 &   8 &  30 &  0 \\
117+097-087 & 03 34 07.28 & -58 24 46.6 & 16.11 & 17774 & 15.19 $\pm$ 0.08 &  1.09 $\pm$ 0.10 & -0.62 $\pm$ 0.24 &  23 &  10 &  96 &  0 \\
144+044-118 & 20 47 21.43 & -57 52 35.6 & 15.32 &  3233 & 14.44 $\pm$ 0.13 &  1.09 $\pm$ 0.17 & -0.52 $\pm$ 2.25 &  35 &  26 & 164 &  0 \\
144-126-031 & 21 12 03.18 & -59 26 26.9 & 16.24 &  9493 & 14.63 $\pm$ 0.08 &  1.33 $\pm$ 0.13 &  0.00 $\pm$ 0.00 &  17 &  15 & 139 &  0 \\
144+095-133 & 20 40 26.42 & -57 34 32.3 & 16.93 & 10964 & 15.27 $\pm$ 0.01 &  1.56 $\pm$ 0.02 & -0.41 $\pm$ 0.10 &  16 &  11 & 129 &  0 \\
144+091+082 & 20 39 21.60 & -61 34 36.5 & 15.69 & 22306 & 14.66 $\pm$ 0.34 &  1.65 $\pm$ 0.45 &  0.00 $\pm$ 0.00 &  36 &  25 &  87 &  2 \\
145+055+050 & 21 23 44.04 & -61 02 29.3 & 15.32 &  4404 & 12.96 $\pm$ 0.01 &  1.07 $\pm$ 0.01 & -3.01 $\pm$ 0.01 &  38 &  12 &  80 &  3 \\
145-099-026 & 21 46 53.28 & -59 35 26.3 & 16.36 &  8053 & 14.77 $\pm$ 0.10 &  1.60 $\pm$ 0.14 & -0.25 $\pm$ 0.66 &  20 &  15 &  53 &  0 \\
145+035+005 & 21 26 58.51 & -60 13 18.9 & 15.07 &  8660 & 13.19 $\pm$ 0.12 &  1.35 $\pm$ 0.18 & -0.75 $\pm$ 0.22 &  54 &  40 & 129 &  2 \\
147-070+099 & 22 59 03.11 & -61 53 25.3 & 17.02 &  7787 & 16.25 $\pm$ 0.36 &  1.04 $\pm$ 0.38 & -1.43 $\pm$ 0.69 &  24 &  17 &  93 &  2 \\
148+099-068 & 23 11 40.26 & -58 46 11.3 & 16.55 &  3376 & 15.48 $\pm$ 0.03 &  0.71 $\pm$ 0.03 &  0.00 $\pm$ 0.00 &  20 &  11 &  24 &  0 \\
149-013-101 & 00 01 43.21 & -53 11 47.1 & 14.60 &  9773 & 12.86 $\pm$ 0.07 &  1.68 $\pm$ 0.12 &  0.00 $\pm$ 0.00 &  38 &  30 & 167 &  0 \\
149+031+040 & 23 55 51.03 & -55 48 43.0 & 16.27 &  9477 & 14.78 $\pm$ 0.01 &  1.51 $\pm$ 0.01 & -0.22 $\pm$ 0.06 &  27 &   7 &   9 &  0 \\

   \hline
  \end{tabular}
 \end{tt}
 \end{minipage}
\end{table*}

\end{document}